\documentclass[12pt]{article}
\usepackage{graphicx}
\usepackage{epstopdf}
\usepackage{changes}
\usepackage{hyperref}

\newlength{\vshift}
\input epsf.tex
\newlength{\hshift}

\usepackage{amssymb}
\usepackage{amsmath}
\usepackage{amsthm}
\usepackage{amsopn}
\usepackage{pifont}
\usepackage{xcolor}
\usepackage{cite}
\usepackage{bm}

\def\uno{\mbox{1 \kern-.59em {\rm l}}}

\def\be{\begin{equation}}
\def\ee{\end{equation}}
\def\ba{\begin{eqnarray}}
\def\ea{\end{eqnarray}}
\def\o{\overline}
\def\lo{\longrightarrow}

\def\hh{\hskip 2cm}
\def\la{\langle}
\def\ra{\rangle}
\def\a{\alpha}
\def\b{\beta}

\newcommand{\ket}[1]{\vert #1 \rangle}
\newcommand{\bra}[1]{\langle #1 \vert}

\newcommand{\braket}[2]{\langle #1 \vert #2 \rangle}

\DeclareMathOperator{\Tr}{\mathrm{Tr}}
\newcommand\M{min(k,k')}

\begin{document}

 \vspace*{3cm}

\begin{center}

{\bf{\ Creating Maximally Entangled States by Gluing }}

\vskip 4em

{\bf  Zahra ~Raissi \footnote{e-mail: z\underline{ }raissi@physics.sharif.edu} \: and
\:{\bf Vahid ~Karimipour, \footnote{e-mail: vahid@sharif.edu}}}

\vskip 1em
{Department of Physics, Sharif University of Technology,

P.O. Box 111555-9161,

Tehran, Iran.}
\end{center}

\vspace*{1.9cm}

\begin{abstract}

We introduce a general method of gluing multi-partite states and show that entanglement swapping is a special class of a wider range of gluing operations. The gluing operation of two $m$ and $n$ qudit states consists of an entangling operation on two given qudits of the the two states followed by operations of measurements of the two qudits in the computational basis. Depending on how many qudits (two, one or zero) we measure, we have three classes of gluing operation, resulting respectively in $m+n-2$, $m+n-1$ or $m+n$ qudit states.  Entanglement swapping belongs to the first class and has been widely studied, while the other two classes are presented and studied here. In particular we study how larger GHZ and W states can be constructed when we glue the smaller GHZ and W states by the second method. Finally we prove that when we glue two states by the third method, the $k$-uniformity of the states is preserved. That is when a $k$-uniform state of $m$ qudits is glued to a $k'$-uniform state of $n$ qudits, the resulting state will be a $\M$-uniform of $m+n$ qudits.  

\end{abstract}


\section{Introduction}

The fundamental role that multi-partite entangled states play in many quantum information processing tasks, 
 \cite{Raussendorf, Aolita, Barreiro}, like measurement-based quantum computing, \cite{Raussendorf,Benjamin,Briegel},  quantum error correction \cite{Scott,Calderbank,Steane, Cleve}, quantum secret sharing \cite{Hillery} and multi-party teleportation \cite{Karlsson, Dur-teleportation}, and finally quantum networks \cite{Cirac-network, Kimble }, has resulted in a rather intensive study of their characterization \cite{Facchi2008,Facchi2009,Arnau,Arnaud,Goyenche}, and their experimental realization \cite{Jennewein-2001,Sciarrino-2002,Jennewein-2005,Ma-2012,Su}.   
Despite all this progress, and in contrast to the case of bi-partite entanglement, our knowledge about multi-partite entanglement is still in its infancy. For example, although there is a well-defined order in the entanglement of bi-partite states, it is now known that for 3-party qubit states such an order is not possible, as there are two LOCC inequivalent classes \cite{Dur} and as the number of parties grow the number of these classes will also grow. \\ 

Among the multi-partite states, a very special class has now attracted a much higher attention is the class of $k$-uniform states  \cite{Arnaud}. These are the states which have the property that all of their reductions to $k$ parts are maximally mixed. As an example,  the Greenberger-Horne-Zeilinger (GHZ) state
$|GHZ\ra=\frac{1}{\sqrt{2}}(|0000\ra+|1111\ra)$ \cite{Greenberger} is a 1-uniform state but not a 2-uniform state of qubits. On the other hand the W state $|W\ra=\frac{1}{\sqrt{3}}(|100\ra+|010\ra+|001\ra)$ is not even a 1-uniform state. Obviously a $k$-uniform state is an $l$-uniform state for $l<k$.  Furthermore the Schmidt decomposition shows that a state can be at most $[\frac{n}{2}]$-uniform, where $[x] $ denotes the integer part of $x$.  Particularly interesting are those $n$-qubit states which are $[\frac{n}{2}]$-uniform. Such states are called Maximally Multi-partite Entangled States or MMES for short. In some works \cite{Arnau,Goyenche} these states are also called Absolutely Maximally Entangled (AME) states. \\

It should be noted that in contrast to the two-partite case,  it is not in general possible to  order multipartite pure entangled state by a single quantity. In fact there is even no consensus as to which characteristic may define maximal entanglement even for symmetric states. For example in \cite{GisinBechmann-Pasquinucci} different criteria like maximal violation of multi-partite Bell inequalities, maximal fragility of the states, maximality of the mutual information of measurement outcomes, when a subset of qubits are measured and maximal mixedness of partial subsystems are   studied in detail.  Their  intriguing conclusion is that the last criterion is in fact not a good measure for characterizing multi-partite entanglement, as the other ones. However as stated in the introduction, maximal mixedness of subsystems is important for some important quantum communication tasks, for example in parallel teleportation and quantum secret sharing with arbitrary threshold access structures \cite{Arnau}.  Therefore it is justified that a great deal of attention has been paid to states having this property. Besides these  applications,  mathematical characterization of such states has revealed nice connections \cite {Goyenche} with branches of combinatorics like Latin squares, orthogonal arrays,  combinatorial designs and Hadamard matrices. \\ 

In view of the large number of constraints that an AME state should satisfy, it is obvious that the existence, let alone the systematic construction of these states is a highly nontrivial problem. 
In fact it has been shown that these states exist only for special values of $n$ (the number of qubits). For instance there is no AME state for four-qubits \cite{Gour} although there are AME state of five and six-qubits \cite{Laflamme,Borras}. 
The existence of an AME states of seven-qubit was an open question until very recently where it was shown \cite{Huber}, there is no such state. It had already been shown that no  AME state exists  for eight or more qubits \cite{Facchi2008,Facchi2009}. We should stress that these results are specific to qubits and for any $n$, there are AME state if we choose the dimension of each part $d$ large enough \cite{Arnau}. In view of the fact that such states may not always exist, one can loose the criteria for maximal mixedness for all bi-partitions and resort to a measure like 
 average purity of entanglement over all bi-partitions  \cite{Scott,Facchi2008}, which is defined as follows: 
 
 \ba
\pi_{ME}:=\binom{n}{k}^{-1} \sum_{i} \pi_{i}, \label{potential}
\ea
where $k=[n/2]$ and $\pi_{i}=Tr _{i} \rho_{i} ^{2}$ is the purity of an $[n/2]$-qubit subsystem $k$. It is obvious that for qubit cases $\frac{1}{2^{[n/2]}}\leq\pi_{ME}\leq1$. The state is fully factorized if $\pi_{ME}=1$, and maximally entangled if $\pi_{ME}=\frac{1}{2^{[n/2]}}$.  
The maximally or nearly maximally entangled states are those which minimize $\pi_{ME}$ which is the average purity over all bipartite partitions. This property allows to recast the problem of finding such states to optimization problems in statistical mechanics, therefore establishing a fruitful link between the two fields \cite{Facchi2008,Scott}.\\


In view of their conceptual and practical importance, it is highly desirable to investigate whether  or not $k$-uniform states can be prepared by some simple method starting from bi-partite states like the Bell states. Since preparation of Bell states  and also implementation of two-qubit entangling gates, seem within the reach of the present technology, application of the gluing procedure proposed in this paper may lead to construction of larger $k$-uniform states, starting from smaller ones. \\

To put the gluing method into a proper context, we compare it with entanglement swapping and show that it naturally falls into this class of operations. In entanglement swapping two nodes of states $|\Psi\ra_n$ and $\Phi\ra_m$, where $n$ and $m$ are the number of parts (i.e. qubits) of these states are measured in the Bell basis. The rest of the two states then projects onto a highly entangled states of $m+n-2$ parts. However, one envisage a Bell measurement as an entangling operation followed by two single qubit measurements on the two nodes (i.e. in the computational basis for qubits). Therefore one may ask what will be the result of this operation if after the entangling operation we measure only one of the qubits or even none of them, where respectively $m+n-1$ and $m+n$ qubits remain at the end. Looking at the entanglement swapping in this way, we see that there are three types of gluing operations depending on whether we measure two, one or no qubits. The first one is the standard entanglement swapping and the next two are introduced in this paper for the first time. \\

Using these new gluing methods, we show how n-partite GHZ states can be constructed once 3-quibt GHZ states are available or how large asymmetric W states can be glued from smaller ones. More importantly we show how  an $m$-qudit $k$-uniform state can be glued to an $n$-qudit $k'$-uniform state, resulting in an $m+n$-qudit state which is $\M$-uniform. Proof of the $\M$ uniformity of this state is non-trivial and will be presented as a theorem. \\

As far as the experimental realization of this gluing is concerned, since its main ingredients are application of two qubit gates and single qubit measurements, it is as feasible as entanglement swapping which has recently been achieved for multi-partite states by several groups \cite{Jennewein-2001,Sciarrino-2002,Jennewein-2005,Ma-2012,Su}. \\

This paper is organized as follows.  In section \ref{sec:General-considerations-on-gluing-of-multi-qubit-states}, we discuss different types of gluing and explain their differences. In three subsections of this section we explain each method in detail and give examples. Specifically in the third subsection, we prove our theorem on   gluing uniform states.  
The paper ends with a conclusion and outlook.

\section{Different types of gluing}\label{sec:General-considerations-on-gluing-of-multi-qubit-states}

\begin{figure}[t]
\center
\includegraphics[width=0.55\columnwidth]{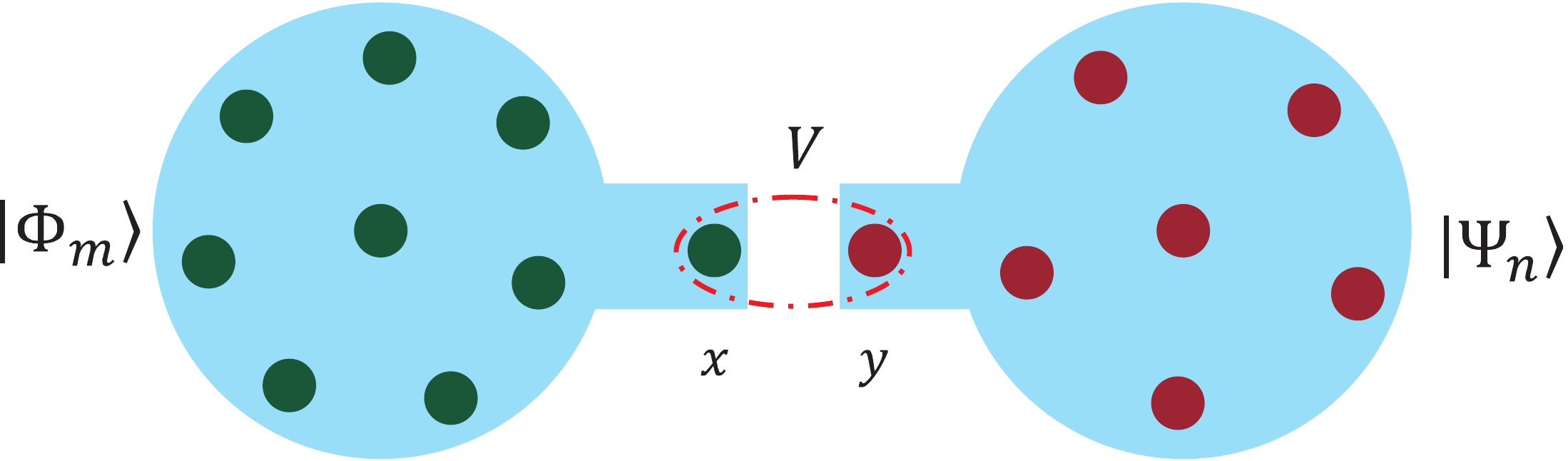}\\
\setlength{\abovecaptionskip}{61pt plus 5pt minus 1pt}
\caption{A graphic representation of the arrangement of two initial entangled pure states $|\Phi\ra_m$ and $|\Psi\ra_n$, such that two of the qubits $x$ and $y$ are possessed by one common lab. For gluing this states we perform on $x$ and $y$, the two qubit unitary operator $V$ and then measure the qubits in the computational basis.}
\label{fig:two-general-states-one-common-lab}
\end{figure}

In this section, we introduce the gluing operation and put it into a proper context where the entanglement swapping appears to be a special case. In fact, entanglement swapping is the earliest gluing method where two pairs of Bell states are joined to each other by making a Bell measurement on the two qubits of each pair, to make a Bell state at a larger distance, figure (\ref{fig:two-general-states-one-common-lab}). A  Bell measurement can be thought of a  suitable entangling operation followed by a measurement in the computational basis. 
In general, one can take an $m$-qudit state $|\Phi\ra$ and an $n$-qudit  $|\Psi\ra$ and then glue them along two of their qudits by entangling these two qudits.  We can then ask what happens if we do the entangling operation (which does the actual act of gluing) but then follow it either by no measurement or by measurement of only one qudit.  This means that depending on how many qubits we measure after the entangling operation, we can have three different types of gluing operation.\\

{\bf Definition:} Let $V$ be an entangling two-qudit gate $V:=\sum_{i,j}|\chi_{i,j}\ra\la i,j|$, where $\{|\chi_{i,j}\ra\}$ a basis of maximally entangled states and let $x$ and $y$ be two specific qudits of the states $|\Phi\ra$ and $|\Psi\ra$, respectively. For gluing the two states along the two qudits, we perform the gate $V$ on $x$ and $y$  and then measure either both, or only one or none of the qudits $x$ and $y$ in the computational basis. The resulting state will be denoted by  
\be\label{general-gluing-3different}
|\Phi\ra\diamond_{\star\star}|\Psi\ra,\qquad |\Phi\ra\diamond_{\star}|\Psi\ra,\qquad |\Phi\ra\diamond|\Psi\ra, 
\ee 
where the number of $\star$ `s denote the number of qudits which are measured in the computational basis. When gluing two $m$- and $n$-qudit states by these methods, then from left to right in (\ref{general-gluing-3different}), the glued state will contain 
$m+n-2,\ m+n-1$ and $m+n$ qudits respectively. 
In this notation $|\Phi\ra\diamond_{\star\star}|\Psi\ra$ denotes the usual entanglement swapping and the other two operations are the ones that we will study in this paper.\\

‎‎‎‎

\subsection{Gluing by double-qudit measurement: Entanglement swapping}\label{sec:Entanglement-swapping}

 For simplicity we restrict our discussion to qubits, although everything can be generalized to the case of qudits. Let $|\Phi\ra$ and $|\Psi\ra$ be two $m$- and $n$-qubit states which we expand as follows:
\be\label{extended-one-qubit-Phi-and-Psi}
|\Phi\ra=|\phi_0\ra_{\o{x}}|0\ra_x + |\phi_1\ra_{\o{x}}|1\ra_x ,\qquad |\Psi\ra=|0\ra_y|\psi_0\ra_{\o{y}}+|1\ra_y|\psi_1\ra_{\o{y}},  
\ee
where $x$ and $y$ are the qubits which we want to glue and $\o{x}$ and $\o{y}$ are the rest of qubits in $\ket\Phi$ and $\ket\Psi$ respectively. Now consider the two qubit gate
\be\label{entangling-gate-qubit}
V=\frac{1}{\sqrt{2}}\left(\begin{array}{cccc} 1 & & & 1 \\  & 1 & 1 & \\ & 1 & -1 & \\ 1 & & & -1 \end{array}\right).
\ee
If we act on the two qubits $x$ and $y$ by this gate and then measure these qubits in the computational basis, the rest of the state will project onto the following, depending on outcome of the measurement: 

\ba
|00\ra_{x,y}\qquad &\lo& \qquad \frac{1}{\sqrt{2}}\left(|\phi_0\ra|\psi_0\ra+|\phi_1\ra|\psi_1\ra\right)\cr
|01\ra_{x,y}\qquad&\lo& \qquad \frac{1}{\sqrt{2}}\left(|\phi_0\ra|\psi_1\ra+|\phi_1\ra|\psi_0\ra\right)\cr
|10\ra_{x,y}\qquad &\lo& \qquad \frac{1}{\sqrt{2}}\left(|\phi_0\ra|\psi_1\ra-|\phi_1\ra|\psi_0\ra\right)\cr
|11\ra_{x,y}\qquad&\lo& \qquad\frac{1}{\sqrt{2}}\left(|\phi_0\ra|\psi_0\ra-|\phi_1\ra|\psi_1\ra\right).
\ea
or more compactly as
\be
|m,n\ra_{x,y}\qquad\lo \qquad \frac{1}{\sqrt{2}}\sum_j(-1)^{mj}|\phi_j\ra|\psi_{j+m+n}\ra.
\ee
This operation of entanglement swapping is symbollically written as 
\be
|\Xi\ra_{m+n-2}=|\Phi\ra_{m}\diamond_{\star\star} |\Psi\ra_n.\label{twostar}
 \ee
As more concrete examples, one can easily show that 
\be
|GHZ_n\ra \diamond_{\star\star}|GHZ\ra_m = |GHZ\ra_{m+n-2}
\ee
This of course shows that $\star\star$-gluing cannot be used to produce $|GHZ\ra_3$ from Bell states, however if we produce $|GHZ\ra_3$ by some other method we can use (\ref{twostar}) to produce all other $|GHZ\ra_n$ states with $n>3$. \\

The same is also true for $|W\ra_n$ states, where $|W\ra_3=\frac{1}{\sqrt{3}}(|100\ra+|010\ra+|001\ra)$ and so forth. A $|W\ra_n$ state can be written as
\be
|W\ra_n=\frac{1}{\sqrt{n}}\left(\sqrt{n-1}|W\ra_{n-1}|0\ra_x+|0^{n-1}\ra|1\ra_x\right)
\ee
similarly
\be
|W\ra_m=\frac{1}{\sqrt{m}}\left(\sqrt{m-1}|0\ra_{y}|W\ra_{m-1}+|1\ra_y|0^{m-1}\ra\right).
\ee
Performing a Bell measurement on $x$ and $y$, we obtain with probability $\frac{m+n-2}{mn}$ the results $|\psi^{\pm}\ra_{xy}=\frac{1}{\sqrt{2}}(|01\ra\pm |10\ra)$ and the rest of the state will project to a state which is equivalent to
\be
|W\ra_{n+m-2}\propto \left(|W\ra_{n-1}|0^{m-1}\ra+|0^{n-1}\ra|W\ra_{m-1}\right),
\ee
thus we have
\be
|W\ra_n\diamond_{\star\star} |W\ra_m=|W\ra_{m+n-2}.\label{gluing_w}
\ee
Again this shows that $|W\ra_3$ state  cannot be produced by joining Bell states, but once $|W\ra_3$ is at hand, we can use it to produce all the other $|W\ra_n$ states via (\ref{twostar}) one by one. \\

In this paper, we focus on the other two types of gluing, namely those which use  one-qubit measurement $|\Phi\ra\diamond_{\star} |\Psi\ra$ and no measurement $|\Phi\ra\diamond|\Psi\ra$.

\subsection{Gluing by single qudit measurement}\label{sec:Gluing-by-single-qudit-measurement}

Let us now see how this gluing operation can be repeated in a systematic way. This procedure is specially useful when we consider a linear chain of laboratories which try to construct a multi-partite entangled state or a quantum network, since as shown above and detailed below, in this type of gluing, the number of qudits increases at each stage of gluing.
For simplicity we restrict ourselves to the qubit case, although generalization to the qudit case is straightforward.
So, consider  the state (\ref{extended-one-qubit-Phi-and-Psi})
\be \label{extended-one-qubit-Phi}
|\Phi\ra=|\phi_0\ra|0\ra_x+|\phi_1\ra|1\ra_x,
\ee
where we are to glue the qubit $x$ to a Bell state $|\phi^+\ra_{yz}$ along the qubits $x$ and $y$. Here after the entangling gate $V$ in (\ref{entangling-gate-qubit}, we measure only the qubit $x$. 
Inserting the state $|\phi^+\ra_{yz}$ on the right hand side of $|\Phi\ra$ and acting by $V$ on $x$ and $y$, we find
\ba
(V)_{xy}|\Phi\ra|\phi^+\ra_{yz}&=&|\phi_0\ra|\phi^+\ra_{xy}|0\ra_z+|\phi_0\ra|\psi^+\ra_{xy}|1\ra_z\cr &+&|\phi_1\ra|\psi^-\ra_{xy}|0\ra_z+|\phi_1\ra|\phi^-\ra_{xy}|1\ra_z,\label{actv}
\ea
which can be rewritten in the form:
\ba
(V)_{xy}|\Phi\ra|\phi^+\ra_{yz}&=&\left[ |\phi_0\ra|\phi^+\ra_{yz}+|\phi_1\ra|\psi^+\ra_{yz}\right]|0\ra_x\cr &+&\left[ |\phi_0\ra|\psi^+\ra_{yz}-|\phi_1\ra|\phi^+\ra_{yz}\right]|1\ra_x.
\ea
Therefore by measuring the qubit $x$, we find two different results for the remaining state, depending on the outcome of the measurement.   The resulting state in case of a $0$ result on $x$ is  given by
\be
|\Phi\ra=|\phi_0\ra|\phi^+\ra_{yz}+|\phi_1\ra|\psi^+\ra_{yz},
\ee
and in case of a $1$ result on $x$ is given by
\be
|\Phi'\ra=|\phi_0\ra|\psi^+\ra_{yz}-|\phi_1\ra|\phi^+\ra_{yz}.
\ee
However these two states are easily converted to each other by a local operation, namely:
\be
|\Phi\ra=Z_{y}\otimes (ZX)_z |\Phi'\ra.
\ee
This means that the process leads deterministically to one of the above states. In order to repeat the process, we consider only the first state, and write the state as $|\Phi^{(1)}\ra$, (to indicate that it is the result of one single gluing) in a way which resembles the initial state given in (\ref{extended-one-qubit-Phi}), namely
\be
|\Phi^{(1)}\ra=|\phi^{(1)}_0\ra|0\ra_z+|\phi^{(1)}_1\ra|1\ra_{z}, \label{expand-chain}
\ee
where
\be
(|\phi^{(1)}_0\ra,\ |\phi^{(1)}_1\ra) = \frac{1}{\sqrt{2}}(|\phi_0\ra,\ |\phi_1\ra)\left(\begin{array}{cc}|0\ra & |1\ra \\ |1\ra & |0\ra\end{array}\right),\label{define-g}
\ee
or compactly as
\be
(|\phi^{(1)}_0\ra,\ |\phi^{(1)}_1\ra) = (|\phi_0\ra,\ |\phi_1\ra){\cal G}_1,
\ee
where
\be
{\cal G}_1=\frac{1}{\sqrt{2}}\left(\begin{array}{cc}|0\ra & |1\ra \\ |1\ra & |0\ra\end{array}\right).
\ee

After $n$ steps of gluing we find 
$
(|\phi^{(n)}_0\ra,\ |\phi^{(n)}_1\ra) = (|\phi_0\ra,\ |\phi_1\ra){{\cal G}_1}^n,\label{define-gn}
$
where the matrix ${{\cal G}_1}^n$  is given by
\be
{{\cal G}_1}^n = \frac{1}{\sqrt{2^{n-1}}}\left(\begin{array}{cc} |e_n\ra & |o_n\ra \\ |o_n\ra & |e_n\ra\end{array}\right),
\ee
in which  $|e_n\ra$ and $|o_n\ra$ are respectively the normalized even and odd $n$-qubit states, i.e. (for n=3)
\ba
|e_3\ra&=&\frac{1}{2}\left(|000\ra+|011\ra+|101\ra+|110\ra\right)\cr
|o_3\ra&=&\frac{1}{2}\left(|111\ra+|100\ra+|010\ra+|001\ra\right).
\ea
If we start with an initial two qubit state  $|\Phi\ra=|\phi^+\ra$, then after gluing of $n$ Bell states, we obtain the $n+1$-qubit state which is a uniform superposition of all $n+1$ qubit states with even parity. This state belongs to the 1-uniform class. \\

Instead of the operator $V$, eq. (\ref{entangling-gate-qubit}) we can use many other two-qubit gates for gluing and in each case we obtain two different outputs after measurement of the qubit $x$, each projecting the rest of qubits to a different state depending on the outcome. In fact once the gluing operator is chosen, the recursion matrix is determined uniquely. One can even use this correspondence to choose the gluing operator on the basis of recursion matrix which is demanded. To see this we repeat the calculation in (\ref{actv}), for a general two-qubit gate $V$.
\ba\nonumber
V_{xy}|\Phi\ra|\phi^+\ra_{yz}&=& \left(|\phi_0\ra V_{0j,00}|j\ra + |\phi_1\ra V_{0j,10}|j\ra\right)|0\ra_z \nonumber\\
&+&  \left(|\phi_0\ra V_{0j,01}|j\ra + |\phi_1\ra V_{0j,11}|j\ra\right)|1\ra_z \label{general-act-v}
\ea
where a summation over repeated dummy indices is understood. Therefore for a general gluing operator and (for the outcome $x=0$)
we find the recursion matrix
\be
{\cal G}= \left(\begin{array}{cc} V_{0j,00}|j\ra & V_{0j,01}|j\ra \\ V_{0j,10}|j\ra& V_{0j,11}|j\ra\end{array}\right).\label{general-recursion}
\ee
This means that we can choose our gluing operator on the basis of the recursion matrix that we want. We can choose those gluing operators for which the two resulting states can be connected with local actions of unitary operators. In this case the result of the gluing is deterministic and we proceed only with one of the states. Note that not all gluing operators have this property. In the following we introduce a few other operators and their corresponding recursion matrices which will be used in the sequel.
The method of calculation in all these cases are similar and will not be repeated, so we only list the gluing operators and the corresponding recursion matrices. Three other entangling operator and their corresponding recursion matrices, which will be used below are as follows:

\ba
V_2&=&\frac{1}{\sqrt{2}}\left(\begin{array}{cccc}1 & 0 & 0 & 1 \\ 0 & 1 & -1 & 0 \\ 0 & 1 & 1 & 0 \\ 1 & 0 & 0 & -1\end{array}\right)\hh {\cal G}_2=\left(\begin{array}{cc}|0\ra & |1\ra \\ -|1\ra & |0\ra\end{array}\right)\label{v2}\cr
V_3&=&\left(\begin{array}{cccc}
1 & 0 & 0 & 0 \\ 0 & 0 & 0 & 1 \\ 0 & 0 & 1 & 0 \\  0& 1 & 0 &0
\end{array}\right)\hh \ \ \ \  \ \ \ \ \ \ {\cal G}_3=\left(\begin{array}{cc}|0\ra &  \\  & |1\ra\end{array}\right)\label{v3}\cr
V_4&=&\left(\begin{array}{cccc}
\frac{1}{\sqrt{2}} & 0 & 0 & \frac{1}{\sqrt{2}} \\ 0 & 1 & 0 & 0 \\ 0 & 0 & 1 & 0 \\  \frac{1}{\sqrt{2}}& 0 & 0 &-\frac{1}{\sqrt{2}}
\end{array}\right)\hh\ \  \ \ {\cal G}_4=\left(\begin{array}{cc}\frac{1}{\sqrt{2}}|0\ra & |1\ra \\ 0 & \frac{1}{\sqrt{2}}|0\ra\end{array}\right)\label{v4}
\ea

{\bf Remark:}
If $|\phi_0\ra$ and $|\phi_1\ra$ can be  converted to each other by using local operations then, we can convert the two result states, for the outcome $x=0$ and $x=1$, by using local operations. Otherwise, the two states which result from different outcomes of measurement of the qubit $x$ cannot be converted to each other by local operations, hence this procedure is probabilistic and we have indicated only the result of $0$ outcome on measurement of $x$. Consequently the state is not normalized and should be normalized only at the end.\\

The toolkit for gluing now consists of recursion relations ${\cal G}_1$, ${\cal G}_2$, ${\cal G}_3$ and ${\cal G}_4$, where one can start from a simple initial state and proceed in various orders to obtain different types of multi-qubit states.
We are now in a position to apply these procedures of gluing to generate various types of states.
We follow the strategy in which laboratories between the two endpoint of the chain do the gluing action consecutively.

\subsubsection{Examples}\label{Examples}
\begin{figure}[t]
\center
\includegraphics[width=0.53\columnwidth]{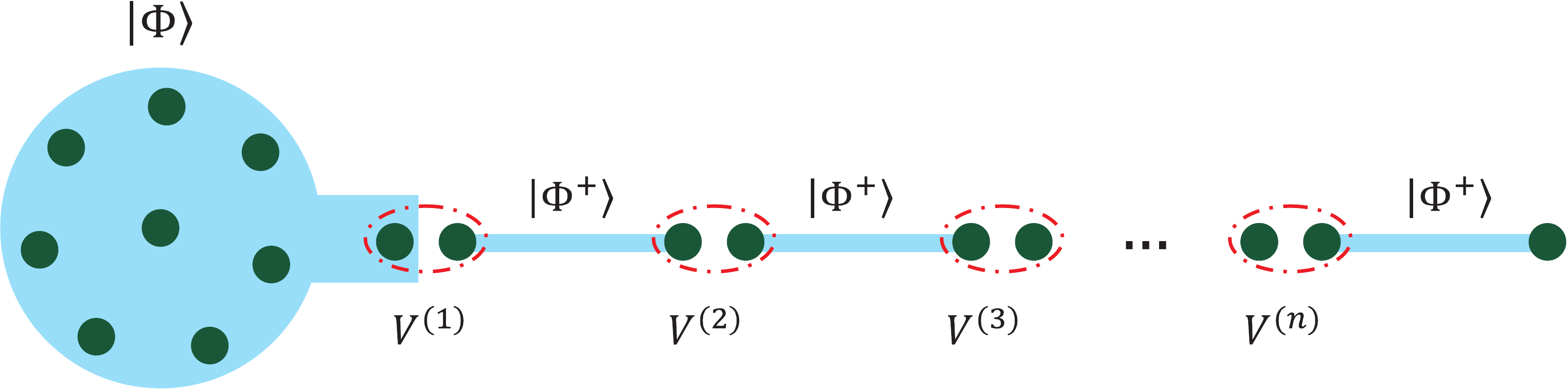}\\
\setlength{\abovecaptionskip}{26pt plus 1pt minus 1pt}
\caption{Gluing Bell states to an initial state $|\Phi\ra$, through linear geometry.}
\label{fig2-Gluing-Bell-states-to-initial-state}
\end{figure}

Consider a situation where laboratories $1$ through $n$, do the action of gluing by operators $V^{(1)}, V^{(2)}, \cdots V^{(n)}$, where the operator $V^{(i)}$ denote the gluing operator used by the $i$th laboratory, as shown in figure (\ref{fig2-Gluing-Bell-states-to-initial-state}). Depending on the number of Bell states which are glued to the initial state (\ref{expand-chain}), and the type of gluing operators various $n+1$ qubit state can be generated.
In the following we present a few interesting examples.\\

{\bf GHZ states:}
The $GHZ_n$ states for all $n$ can be constructed by gluing Bell states with CNOT operators, or equivalently the recursion operator ${\cal G}_3$.  In view of (\ref{general-act-v}), this means that starting with the state $|\phi^+\ra$ for which in the notation of (\ref{extended-one-qubit-Phi}), $|\phi_0\ra=\frac{1}{\sqrt{2}}|0\ra$ and  $|\phi_1\ra=\frac{1}{\sqrt{2}}|1\ra$, we have the relation
 \be
 (|\phi_0^{(n)}\ra,\ |\phi_1^{(n)}\ra)=(\frac{1}{\sqrt{2}}|0\ra,\ \frac{1}{\sqrt{2}}|1\ra)\left(\begin{array}{cc} |00\cdots 0\ra & 0 \\ 0 & |11\cdots 1\ra\end{array}\right),
 \ee
which obviously leads to an $n+1$ qubit GHZ states.\\

{\bf Asymmetric W states:}
Let us now start with the Bell state $|\phi^+\ra$, but use the gluing operator $V_4$ from (\ref{v4}) for which the recursion matrix is ${\cal G}_4$. In this case we obtain the state
\ba
\frac{1}{2}|000\rangle+\frac{1}{\sqrt{2}}|011\rangle+\frac{1}{2}|101\rangle,
\ea
by flipping the last qubit, we obtain the asymmetric $W$ state,
\ba\label{aw}
\frac{1}{2}|001\rangle+\frac{1}{\sqrt{2}}|010\rangle+\frac{1}{2}|100\rangle.
\ea
The asymmetric W state is widely used in quantum information processing, such as perfect teleportation and dense coding \cite{Agrawal}, which can be transformed to the $W$ state by using local approximate protocol \cite{Yildiz}. If the outcome of measurement  is $1$, we get
\ba
\frac{1}{2}|010\rangle+\frac{1}{\sqrt{2}}|100\rangle-\frac{1}{2}|111\rangle,
\ea
which, can be converted to the (\ref{aw}), by using the local operations $X \otimes X \otimes Z$. For the general case, we can use the method (\ref{gluing_w}) to produce all other $|W\ra_n$ states with $n>3$.\\

\begin{figure}[t]
\center
\includegraphics[width=0.42\columnwidth]{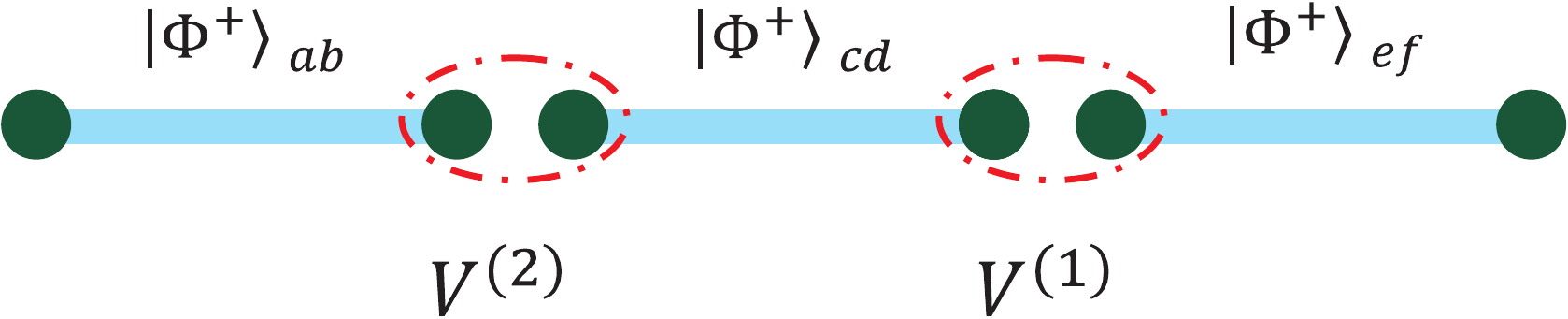}\\
\setlength{\abovecaptionskip}{17pt plus 1pt minus 1pt}
\caption{Gluing operations act on a two-quantum repeater configuration, which consists of three copies of Bell states.}
\label{fig3-Gluing-operations-two-quantum-repeater}
\end{figure}

{\bf A nearly uniform four-qubit state:}
In  case of $n = 4$ qubit systems, it is worth emphasizing that no four-qubit state can have all of its two-qubit reductions maximally mixed \cite{Facchi2008}. In this case and other systems for which do not exist MMES, we seek for the states that their average entanglement are maximal. Let us show how to build this state by gluing in a chain of the Bell pairs. Consider two-repeater configuration such as each of two qubits $bc$ and $de$ of Bell sates $|\phi^+\ra_{ab}$, $|\phi^+\ra_{cd}$, and $|\phi^+\ra_{ef}$ are located in the common labs, (see figure (\ref{fig3-Gluing-operations-two-quantum-repeater})). We glue qubits $bc$ and $de$ by the two qubit gates $V_2$ and $V_1$, respectively, or equivalently the recursion operator ${\cal G}_2$ and ${\cal G}_1$. Then, we obtain the relation
\ba
(|\phi_0^{(2)}\ra,\ |\phi_1^{(2)}\ra)=\frac{1}{\sqrt{2}}(|0\ra,\ |1\ra){\cal G}_2{\cal G}_1=\frac{1}{\sqrt{2}} (|0\ra,\ |1\ra)\left(\begin{array}{cc} |\phi^+\ra & |\psi^+\ra \\ |\psi^-\ra & |\phi^-\ra\end{array}\right)_{ce}.
\ea
The superscript $(2)$ applied on state $|\Phi\ra$ indicate two times of using the action of gluing through a chains of the Bell pairs. This procedure yields the following state of four qubits with the average purity $\pi_{ME}=\frac {1}{3}$.
\ba
|M_4\ra=\frac{1}{2}(|00\ra_{af}|\phi^+ \ra_{ce}+|01\ra_{af}|\psi^ +\ra_{ce}+|10\ra_{af}|\psi^-\ra_{ce}+|11\ra_{af}|\phi^-\ra_{ce}).
\ea

\subsection{Gluing by no qudit measurement}\label{sec:Gluing-no-qudit-measurement}

The thrid kind of gluing is when we do not measure any of the two qudits after entangling them. In this case gluing two $m$- and $n$-qudit states  results in an $(m+n)$-qudit state. This type of gluing has the desirable feature that it produces uniform states when acting on uniform ones. The precise statement is given in the following theorem:  \\

{\bf Theorem:}
Let $|\Phi\ra$ and $|\Psi\ra$ be two $k$- and $k'$-uniform states with $m$ and $n$ qudits respectively. Then the state $|\Phi\ra\diamond|\Psi\ra$ is an $m+n$-qudit state which is $\M$-uniform. \\
Hereafter we assume that $k\leq k'$ and hence we will show that the state $|\Phi\ra\diamond|\Psi\ra$  is a $k$-uniform state. \\

{\bf Proof:}
We remind the reader that a pure state $|\psi\ra$ is $k$-uniform if all the reduced density matrices of its subsystems of less than or equal to $k$ are maximally mixed. Obviously a state which is $k$-uniform will also be $l$-uniform for all $l\leq k$. We can take an arbitrary subset $S$ of the qubits with size $|S|=k$ and expand the state $|\psi\ra$ in terms of a complete orthogonal basis of $S$ in the form 
\be
|\psi\ra=\sum_{\bm{\a}}|\bm{\a}\ra_S\otimes |\psi_{\bm \a}\ra,
\ee
where $\{|\bm{\a}\ra\}$ is an orthonormal basis for the part $S$, and $|\psi_{\bm{\a}}\ra$ are coefficient states of this expansion with support on the complement set of $S$. Since the density matrix of the subset $S$ should be $\frac{1}{d^{|S|}}I$, then we have

\be\label{uniformity-condition}
\la \psi_{\bm \a}|\psi_{\bm \b}\ra=\frac{1}{d^{|S|}}\delta_{{\bm{\a}},{\bm {\b}}} \qquad {\rm if} \qquad |S|\leq k.
\ee

We have to show that this kind of property holds for any subset $S$ of size $k$, no matter how we choose the subset $S$. The difficulty is that in the glued state, this subset can have various forms, it may be contained entirely in the support of $|\Phi\ra$, or the support of $|\Psi\ra$, or it may be split between these two states. Therefore we have to consider various cases. Moreover we have to take into account that this subset may or may not contain the each of the two gluing points ($x$ and $y$) along which the two states have been glued together. \\

\begin{figure}[htb]
\center
\includegraphics[width=0.8\columnwidth]{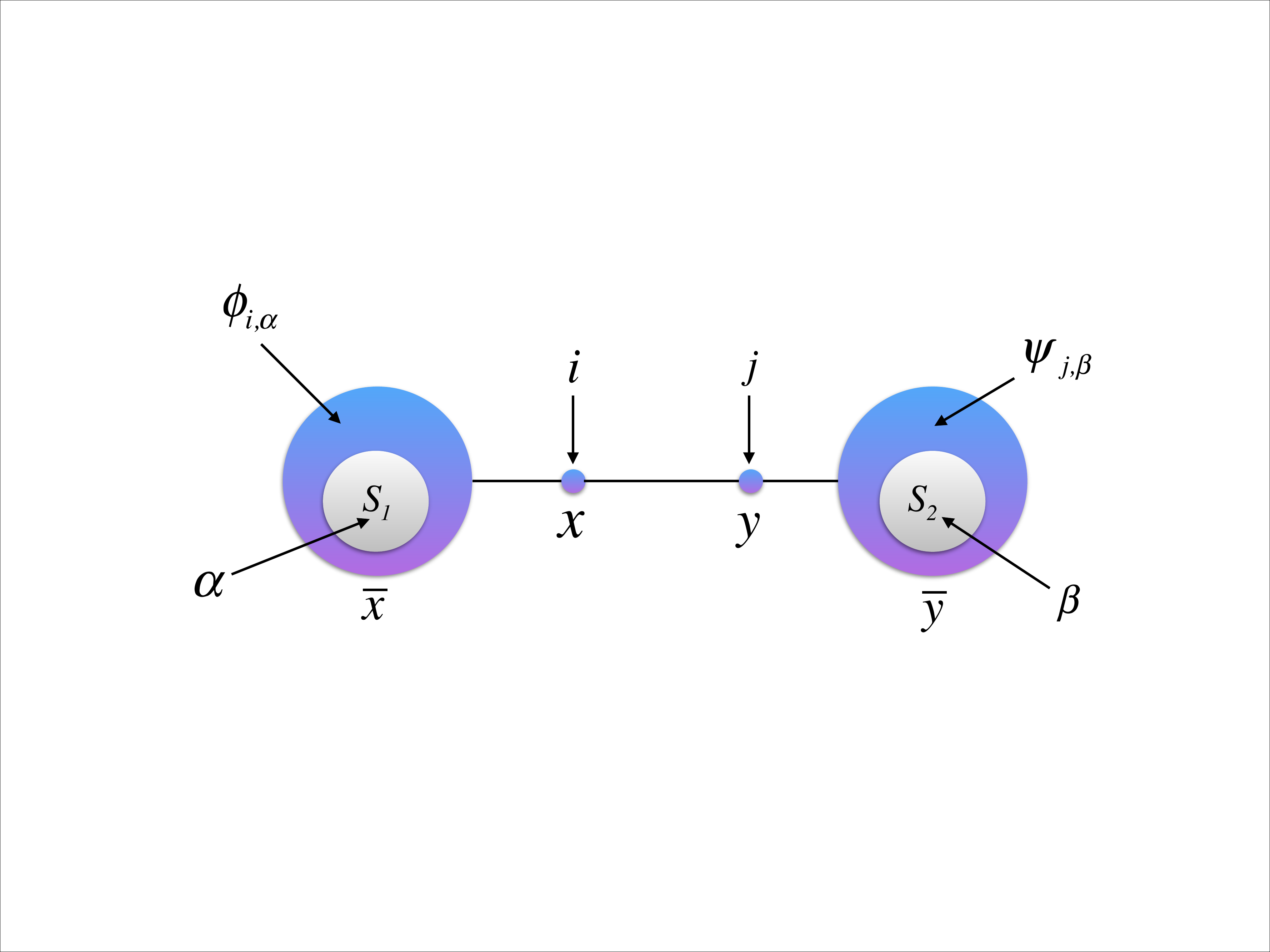}
\setlength{\abovecaptionskip}{61pt plus 5pt minus 1pt}
\caption{Notations and labeling of various parts of the glued states. $x$ and $y$ designate the gluing points. $\o{x}$ denotes the rest of qudits in $\ket{\Phi}$ and $\o{y}$ denotes the rest of qudits in $\ket\Psi$. $S_1$ denotes a subset of qudits in $\o{x}$ and $S_2$ denotes a subset of qudits in $\o{y}$. The full subset of $k$ qudits is equal to $S_1\cup S_2$, $S_1\cup S_2\cup \{x\}$ or $S_1\cup S_2\cup \{x,y\}$ depending on the configurations explained in figure (\ref{fig5-K-UniformGlue}). The total state for this configuration is written as in (\ref{BasicExpand}).}
\label{fig4-K-Uniform-Basic}
\end{figure}

\begin{figure}[htb]
\center
\includegraphics[width=0.9\columnwidth]{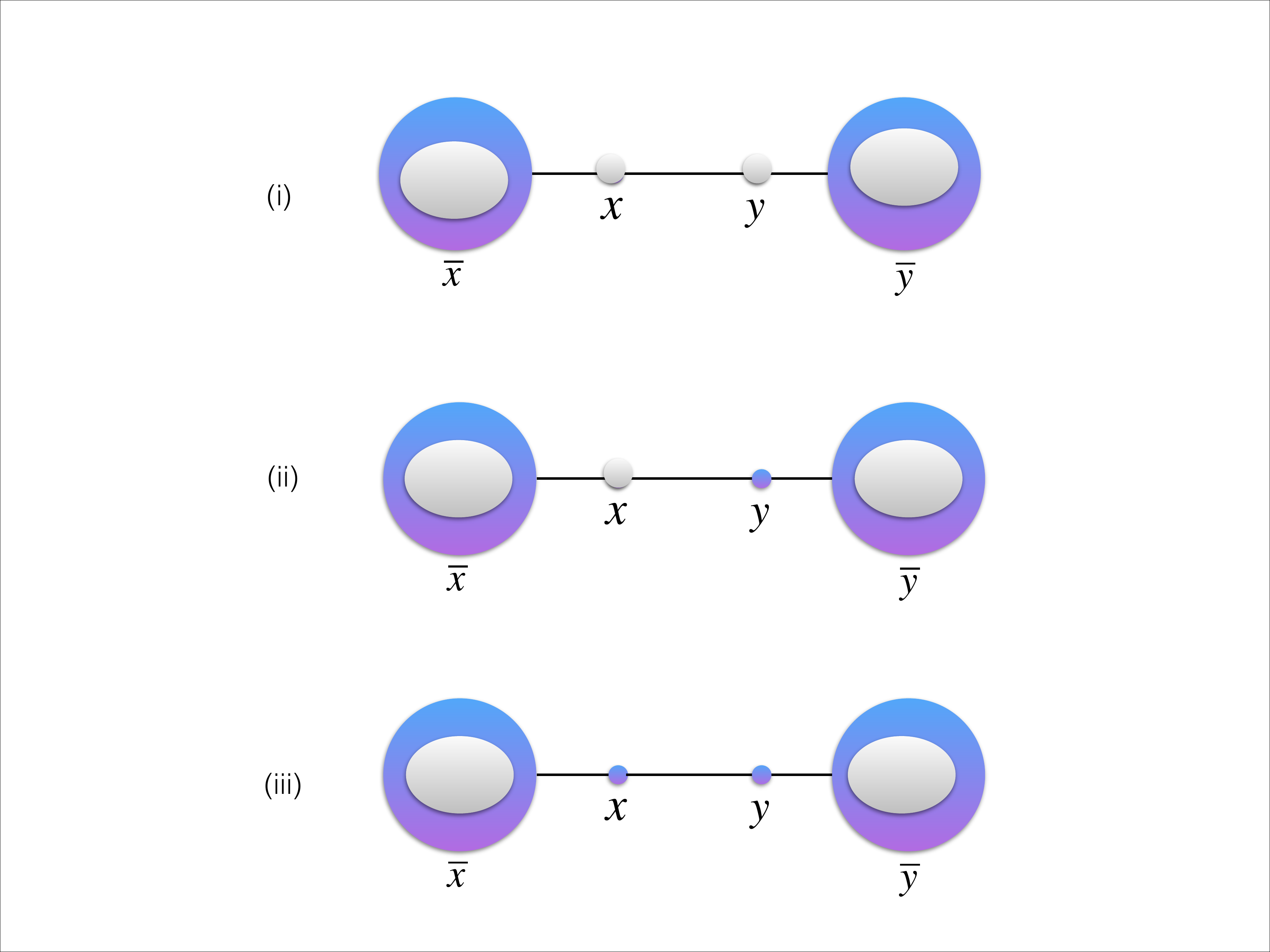}
\setlength{\abovecaptionskip}{61pt plus 5pt minus 1pt}
\caption{
Different configurations for calculating the density matrix of a subset $S$ of $k$-qudits. The gray part depicts the subset $S$ and each of the two subsets in $\o{x}$ or $\o{y}$ can be empty. In (i) the subset $S$ contains both gluing points $x$ and $y$, in (ii), $S$ contains only one gluing point $x$ and in (iii) $S$ contains none of the gluing points. }
\label{fig5-K-UniformGlue}
\end{figure}

When we take a $k$-uniform state and want to glue it along one of its qudits, say $x$, (figure (\ref{fig4-K-Uniform-Basic})), we need to consider an expansion in the form
\be
|\psi\ra=\sum_{i=1}^d\sum_{\bm{\a}} |\psi_{i,\bm{\a}}\ra_{\o{x}/S}   |\bm{\a}\ra_S|i\ra_x.
\ee
For the notations used in this equation please refer to figure (\ref{fig4-K-Uniform-Basic}).
In this case we can use (\ref{uniformity-condition}) and then find the following:

\ba\label{conditions-for-proof}
\la \psi_{i,\bm{\a}}|\psi_{j,\bm{\b}}\ra&=&\frac{1}{d^{|S|+1}}\delta_{i,j}\delta_{\bm{\a},\bm{\b}},\qquad {\rm if} \qquad |S|\leq k-1,\cr
\sum_{i=1}^d\la \psi_{i,\bm{\a}}|\psi_{i,\bm{\b}}\ra&=&\frac{1}{d^{|S|}}\delta_{\bm{\a},\bm{\b}},\qquad\qquad\  {\rm if}\qquad |S|= k.
\ea
To proceed for a proof of the theorem, we start from the explicit expression of the glued state 
\be\label{BasicExpand}
\ket{\Phi}\diamond\ket{\Psi}=\sum_{i,j=1}^d\ket{\phi_i}_{\o x}\ket{\chi_{i,j}}_{x,y}\ket{\psi_i}_{\o y},
\ee
where $\ket{\chi_{i,j}}$ is a basis of 2-qudit maximally entangled state.
We want to show that the reduced density matrix of any subset $S$ of $k$ qudits in this state is maximally mixed. The subset $S$ can have various positions with respect to all the qudits in this large state. Part of $S$ may be in the set $\o{x}$ and part of it may be in the set $\o{y}$ and it may include none, only one or both of the points $x$ and $y$. Without loss of generality, we can consider three different cases, shown in figure (\ref{fig5-K-UniformGlue}). In the following we will consider these cases separately.

‎‎‎‎
 
‎‎‎‎

\begin{itemize}

\item{Case i) The set $S$ contains both of the gluing points, $x$ and $y$, figure (\ref{fig5-K-UniformGlue}-(i))}\\
In this case $S=S_1\cup S_2\cup \{x,y\}$, where $S_1\subset \o{x}$ and $S_2\subset \o{y}$, $|S_1|+|S_2|=k-2$ where either $S_1$ or $S_2$ can be empty. In this case we expand (\ref{BasicExpand}) further as follows: 
\be\label{result-state-of-gluing-no-measure}
\ket{\Phi}\diamond\ket{\Psi}=\sum_{i,j,\bm{\a},\bm{\b}}\ket{\bm{\a}}_{S_1}\ket{\phi_{i,\bm{\a}}}_{\o x/S_1}\ket{\chi_{i,j}}_{x,y}\ket{\bm{\b}}_{S_2}\ket{\psi_{j,\bm{\b}}}_{\o y/S_2}.
\ee
We will then have 
\be
\rho_S=Tr_{\o{x}/S_1}\Tr_{\o{y}/S_2}\left[(\ket{\Phi}\diamond\ket{\Psi})\right(\bra{\Phi}\diamond\ket{\Psi)}].
\ee

Since in this case $|S_1|\ ,\ |S_2| < k$, then we use the first equation in  (\ref{conditions-for-proof}) and directly arrive at the result
\ba
\rho_S&=&\frac{1}{d^{|S_1|+1}}\sum_{\bm{\a}} |\bm{\a}\ra_{S_1}\la \bm{\a}|\otimes \frac{1}{d^{|S_2|+1}}\sum_{\bm{\b}} |\bm{\b}\ra_{S_2}\la \bm{\b}|\otimes \sum_{i,j} |\chi_{ij}\ra_{x,y}\la \chi_{i,j}|\cr
&=& \frac{1}{d^{|S|}}I_{S_1}\otimes I_{S_2}\otimes I_{x,y}= \frac{1} {d^{|S|}}I_S.
\ea
 
\item{Case ii)  The set $S$ contains only one of the gluing points, say $x$, figure (\ref{fig5-K-UniformGlue}-(ii)). }

In this case, we have $S=S_1 \cup S_2 \cup \{x\}$, where $S_1\subset \o{x}$, $S_2\subset o{y}$, and $|S_1|+|S_2|=k-1$. Since in this case both $|S_1|$ and $|S_2|$ are less than $k$, we can use the first equation in (\ref{conditions-for-proof}) and find that:

\ba
\rho_S&=& \Tr_{\o{x}/S_1} \Tr_{\o{y}/S_2} \Tr_{y}\left[\ket{\Phi\diamond\Psi}\bra{\Phi\diamond\Psi}\right]\cr
&=&
\frac{1}{d^{|S_1|+1}}\sum_{\bm{\a}} |\bm{\a}\ra_{S_1}\la \bm{\a}|\otimes \sum_{\bm{\b}}\frac{1}{d^{|S_2|+1}} |\bm{\b}\ra_{S_2}\la \bm{\b}|\otimes \sum_{i,j}\Tr_y |\chi_{ij}\ra_{x,y}\la \chi_{i,j}|
\ea
The last term gives $d\times I_x$ and hence we are left with
\be
\rho_S=\frac{1}{d^{|S_1|+|S_2|+1}}I_{S_1}\otimes I_{S_2}\otimes I_x=\frac{1}{d^k}I_S.
\ee

Finally we come to the third case:

\item{Case iii) The set $S$ contains none of the gluing points, $x$ and $y$, figure (\ref{fig5-K-UniformGlue}-(iii))}\\

In this case $S=S_1\cup S_2$, where $S_1\subset \o{x}$ and $S_2\subset \o{y}$, where either $S_1$ or $S_2$ can be empty and $|S|=k$. We now expand (\ref{BasicExpand}) further as follows: 
\ba
\ket{\Phi}\diamond\ket{\Psi}&=&\sum_{i,j,\bm{\a},\bm{\b}}\ket{\bm{\a}}_{S_1}\ket{\phi_{i,\bm{\a}}}_{\o x/S_1}\ket{\chi_{i,j}}_{x,y}\ket{\bm{\b}}_{S_2}\ket{\psi_{j,\bm{\b}}}_{\o y/S_2}\cr
 &=&\sum_{\bm{\a},\bm{\b}}\ket{\bm{\a}}_{S_1}\ket{\bm{\b}}_{S_2} \ket{\Gamma_{\bm{\a},\bm{\b}}},
\ea
where
\be
\ket{\Gamma_{\bm{\a},\bm{\b}}}=\sum_{i,j}\ket{\phi_{i,\bm{\a}}}_{\o x/S_1}\ket{\chi_{i,j}}_{x,y}\ket{\psi_{j,\bm{\b}}}_{\o y/S_2}.
\ee
To prove $k$-uniformity or maximal mixedness of $\rho_S$, we have to show that 
\be
\braket{\Gamma_{\bm{\gamma},\bm{\delta}}}{\Gamma_{\bm{\a},\bm{\b}}}=\frac{1}{d^k}\delta_{\bm{\gamma},\bm{\delta}}\delta_{\bm{\alpha},\bm{\beta}}.
\ee

To show this we note that in view of the relation $\la \chi_{i,j}|\chi_{k,l}\ra=\delta_{i,k}\delta_{j,l}$, we can write
\ba
\braket{\Gamma_{\bm{\gamma},\bm{\delta}}}{\Gamma_{\bm{\alpha},\bm{\beta}}} &=& 
\sum_{i,j}\braket{\phi_{i,\bm{\gamma}}}{\phi_{i,\bm{\a}}}_{\o x/S_1}\braket{\psi_{j,\bm{\delta}}}{\psi_{j,\bm{\b}}}_{\o y/S_2}.
\ea

We now note that $|S_1|+|S_2|=k$ and hence either both $|S_1|$ and $|S_2|$ are less than $k$ (when none of the two sets $S_1$ and $S_2$ is empty) or one of them is equal to $k$ and the other is zero (when one of the sets is empty). In the first case, we use the second equation in (\ref{conditions-for-proof}) and find that 
\ba
\braket{\Gamma_{\bm{\gamma},\bm{\delta}}}{\Gamma_{\bm{\alpha},\bm{\beta}}} &=& 
\sum_{i,j}\frac{1}{d^{|S_1|+1}}\delta_{\gamma, \alpha}\frac{1}{d^{|S_2|+1}}\delta_{ \delta,\beta}\cr&=&\frac{1}{d^{k}}\delta_{\bm{\gamma},\bm{\delta}}\delta_{\bm{\alpha},\bm{\beta}}.
\ea

In the second case, we use the first equation of (\ref{conditions-for-proof}) and write
\ba
\braket{\Gamma_{\bm{\gamma},\bm{\delta}}}{\Gamma_{\bm{\alpha},\bm{\beta}}} &=& 
\sum_{i}\braket{\phi_{i,\bm{\gamma}}}{\phi_{i,\bm{\a}}}_{\o x/S_1}\sum_j \braket{\psi_{j,\bm{\delta}}}{\psi_{j,\bm{\b}}}_{\o y/S_2}\cr &=& \frac{1}{d^{|S_1|}}\delta_{\bm{\gamma},\bm{\alpha}}\times  \frac{1}{d^{|S_2|}}\delta_{\bm{\delta},\bm{\beta}}=\frac{1}{d^{k}}\delta_{\bm{\gamma},\bm{\alpha}}\delta_{\bm{\delta},\bm{\beta}}.
\ea
\end{itemize}

Note that it is not possible to generate a state with a degree of uniformity higher than $\M$. This is expected since generally by local operation one cannot generate or increase entanglement. 
Therefore we have shown that if $|\Phi\ra$ and $|\Psi\ra$ are two $m$- and $n$-qudit, $k$-uniform and $k'$-uniform states respectively, then $\ket{\Phi}\diamond\ket{ \Psi}$ will be an $m+n$-qudit $\M$-uniform state.

\section{CONCLUSIONS}

We have investigated the general method of gluing multi-partite state together and have shown that entanglement swapping is a special class of a wider range of gluing operations. More precisely we envisage the gluing operation of two $m$ and $n$ qudit states as an entangling operation on two given qudits of the two states followed by operations of measurements of the two qudits in the computational basis. Depending on how many qudits (two, one or zero) we  measure, we have three classes of gluing operation, resulting respectively in $m+n-2$, $m+n-1$ or $m+n$ qudit states.  Entanglement swapping belongs to the first class and has been widely studied, while the other two classes  are presented here. We have shown how these gluing operations can be iterated and how the resulting states can be written in compact form. In particular we have shown how larger GHZ and W states can be constructed when we glue the smaller GHZ and W states by the second method. Finally we have proved that when we glue two states by the last method, the $k$-uniformity of the states is preserved. That is when a $k$-uniform state of $m$ qudits is glued to a $k'$-uniform state of $n$ qudits, the resulting state will be a $\M$-uniform of $m+n$ qudits.

\bibliographystyle{iopart-num}
\providecommand{\newblock}{}

\end{document}